\documentclass[12pt]{article}
\usepackage{graphicx}
\usepackage{endfloat}

\begin{document}
\title{ Experimental implementation of
 generalized Grover's algorithm of multiple marked states and its application}
\author{\small{} Jingfu Zhang$^{1}$ , Zhiheng Lu$^{1}$,Lu Shan$^{2}$,and Zhiwei
Deng$^{2}$  \\
\small{} $^{1}$Department of Physics,\\
\small{}Beijing Normal University, Beijing,
100875, Peoples' Republic of China\\
\small{} $^{2}$Testing and Analytical Center,\\
\small{}  Beijing Normal University,
 Beijing,100875, Peoples' Republic of China}
\date{}
\maketitle
\begin{minipage}{120mm}
\hspace{0.5cm} {\small Generalized Grover's searching algorithm
for the case in which there are multiple marked states is
demonstrated on a nuclear magnetic resonance (NMR) quantum
computer. The entangled basis states (EPR states) are synthesized
using the algorithm.}

 PACS number(s):03.67
\end{minipage}
\vspace{0.3cm}
\section*{1.Introdution}
  Since the quantum searching algorithm was first
proposed by Grover [1], several generalizations of the original
algorithm have been developed [2]-[4]. The generalized algorithm
that we will realize can be posed as follows. Let $N$ basis states
of a system constitute set $D$. A function $F$ is defined as
$F:D\rightarrow \{0,1\}$. The states satisfying $F(x)=1$ are
defined as marked states, which constitute set $M$ with a total
 of $r$ states. The other states in $D$, satisfying $F(x)=0$,
constitute set $\overline{M}$ with a total of $N-r$ states. The
states in $M$ and $\overline{M}$ have amplitudes $k_{i}$ and
$l_{i}$, respectively. A unitary operator $U$ transforms a
predefined basis state $|s>$ into a superposition denoted as
$|g(0)>=U|s>$. $U$ can be almost any valid quantum mechanical
unitary operator. $|g(0)>$ is the initial state for the algorithm.
The phase rotation of marked states is described by
$I_{t}^{\gamma}=\sum_{x} e^{i\gamma F(x)}|x><x|$. Obviously, if
$|x>\in M$, $I_{t}^{\gamma}|x>=e^{i\gamma} |x>$; if
$|x>\in\overline{M}$, $I_{t}^{\gamma}|x>=|x>$. A composite
operator is defined as $G\equiv-UI_{s}^{\beta}U^{\dag }$.
$I_{s}^{\beta}$ is defined as $I_{s}^{\beta}\equiv
I-(1-e^{i\beta})|s><s|$, where $I$ denotes unit matrix. The
problem is to transform $|g(0)>$ to a target state denoted as
$|\psi_{t}>=\sum_{i\in M}k_{i}|i>$ by repeating Grover iteration
$n$ times. When the system lies in $|\psi_{t}>$, measurement
yields $|k_{i}|^{2}$, the probability of the system being in
marked state $|i>$. When $\beta=\pi$, $\gamma=\pi$,
$|s>=|\overline{0}>$, and $U$ is chosen as Walsh-Hadamard (W-H)
transform, the generalized algorithm becomes the original Grover's
algorithm, where $|\overline{0}>$ denotes all qubits in the 0
state.

  E.Biham et al have analyzed generalized Grover's algorithm using
recursion equations [3]. Through introducing an ancilla qubit and
choosing a proper $U$, Grover proposed a theoretical scheme to
synthesize a specified quantum superposition on $N$ states in
$O(\sqrt{N})$ steps using the algorithm [5]. We find that some
special superpositions, such as EPR states, can be synthesized
using the algorithm without the ancilla qubit. Generalized
Grover's algorithm of one marked state has been realized on a
two-qubit NMR quantum computer. G.-L. Long et al realized the
algorithm by choosing the phase rotation as
$I_{t}^{\frac{\pi}{2}}$, while the W-H transform is retained [6].
In our previous work, we realized the algorithm by replacing the
W-H transform by other unitary operator, while the $\pi$ phase
rotation ($I_{t}^{\pi}$) was unaltered [7]. In this paper, we will
realize the generalized algorithm of multiple marked states and
synthesize EPR states. The W-H transform is replaced by other
unitary operator and the phase rotation is chosen as
$I_{t}^{\frac{\pi}{2}}$ or $I_{t}^{-\frac{\pi}{2}}$.

\section*{2.The generalized Grover's algorithm}
  In this section, we will use some results in Ref.[3] to express
the principle of our experiments.

  $n$ applications of $GI_{t}^{\gamma}$ transform $|g(0)>$ into
$|g(n)>$, described by

\begin{equation}\label{1}
  |g(n)>=\sum_{i\in M}k_{i}(n)|i>+\sum_{i\in \overline{M}}l_{i}(n)|i>.
\end{equation}
When $n=0$, the system lies in the initial state

\begin{equation}\label{2}
 |g(0)>=U|s>.
\end{equation}
One can find that $k_{i}(0)=U_{is}$, $i\in M$, and
$l_{i}(0)=U_{is}$, $i\in \overline{M}$, where $U_{is}=<i|U|s>$.
$GI_{t}^{\gamma}$ transforms the amplitudes $k_{j}(n)$, $j\in M$,
to $k_{j}(n+1)=<j|GI_{t}^{\gamma}|g(n)>$ and amplitudes
$l_{j}(n)$, $j\in \overline{M}$ to
$l_{j}(n+1)=<j|GI_{t}^{\gamma}|g(n)>$. The recursion equations
describing such iteration are expressed as

\begin{equation}\label{3}
 k_{j}(n+1)=e^{i\gamma}(1-e^{i\beta})U_{js}\sum_{i\in M}k_{i}(n)U_{is}^{*}
 +(1-e^{i\beta})U_{js}\sum_{i\in \overline{M}}l_{i}(n)U_{is}^{*}-e^{i\gamma}k_{j}(n),
\end{equation}

\begin{equation}\label{4}
 l_{j}(n+1)=e^{i\gamma}(1-e^{i\beta})U_{js}\sum_{i\in M}k_{i}(n)U_{is}^{*}
 +(1-e^{i\beta})U_{js}\sum_{i\in
 \overline{M}}l_{i}(n)U_{is}^{*}-l_{j}(n).
\end{equation}
Without loss of generality, we assume that $U_{is}\neq 0$,
$i=1,2,\cdots$. Valuables $k_{i}^{'}(n)$ and $l_{i}^{'}(n)$ are
defined as

\begin{equation}\label{5}
  k_{i}^{'}(n)=\frac{k_{i}(n)}{U_{is}},
\end{equation}

\begin{equation}\label{6}
  l_{i}^{'}(n)=\frac{l_{i}(n)}{U_{is}}.
\end{equation}
One can easily find that $k_{i}^{'}(0)=1$, $l_{i}^{'}(0)=1$. The
weighted averages are defined as

\begin{equation}\label{7}
  \overline{k^{'}}(n)=\frac{1}{W_{k}}\sum_{i \in M
  }|U_{is}|^{2}k_{i}^{'}(n),
\end{equation}

\begin{equation}\label{8}
  \overline{l^{'}}(n)=\frac{1}{W_{l}} \sum_{i \in
  \overline{M}}|U_{is}|^{2}l_{i}^{'}(n),
\end{equation}
where $W_{k}=\sum_{i\in M}|U_{is}|^{2}$, and $W_{l}=\sum_{i\in
\overline{M}}|U_{is}|^{2}$. With these variables, the recursion
equations can be rewritten as

\begin{equation}\label{9}
 k_{j}^{'}(n+1)=e^{i\gamma}(1-e^{i\beta})W_{k}\overline{k^{'}}(n)
 +(1-e^{i\beta})W_{l}\overline{l^{'}}(n)-e^{i\gamma}k_{j}^{'}(n),
\end{equation}

\begin{equation}\label{10}
 l_{j}^{'}(n+1)=e^{i\gamma}(1-e^{i\beta})W_{k}\overline{k^{'}}(n)
 +(1-e^{i\beta})W_{l}\overline{l^{'}}(n)-l_{j}^{'}(n).
\end{equation}
By averaging over all the marked states in Eq.(9) and over all the
unmarked states in Eq.(10), we find the two recursion equations
for $\overline{k^{'}}(n)$ and $\overline{l^{'}}(n)$ can be
expressed as

\begin{equation}\label{11}
 \overline{k^{'}}(n+1)=e^{i\gamma}(1-e^{i\beta})W_{k}\overline{k^{'}}(n)
 +(1-e^{i\beta})W_{l}\overline{l^{'}}(n)-e^{i\gamma}\overline{k^{'}}(n),
\end{equation}

\begin{equation}\label{12}
\overline{l^{'}}(n+1)=e^{i\gamma}(1-e^{i\beta})W_{k}\overline{k^{'}}(n)
 +(1-e^{i\beta})W_{l}\overline{l^{'}}(n)-\overline{l^{'}}(n).
\end{equation}
Subtracting Eq.(11) from Eq.(9), and Eq.(12) from Eq.(10), one
finds that

\begin{equation}\label{13}
   k_{j}^{'}(n+1)- \overline{k^{'}}(n+1)=-e^{i\gamma}(k_{j}^{'}(n)-
   \overline{k^{'}}(n)),
\end{equation}

\begin{equation}\label{14}
   l_{j}^{'}(n+1)- \overline{l^{'}}(n+1)=-(l_{j}^{'}(n)-
   \overline{l^{'}}(n)).
\end{equation}
Noting that $k_{i}^{'}(0)=1$, $l_{i}^{'}(0)=1$, we find
$\overline{k^{'}}(0)=1$, $\overline{l^{'}}(0)=1$. Using Eqs.(13)
and (14), we obtained

\begin{equation}\label{15}
  k_{i}^{'}(n)= \overline{k^{'}}(n),
\end{equation}

\begin{equation}\label{16}
  l_{i}^{'}(n)= \overline{l^{'}}(n),
\end{equation}
where the subscript $j$ in Eqs.(13) and (14) has been replaced by
$i$.

From the discussion above, for any $U$, $\overline{k^{'}}(n)$ and
$\overline{l^{'}}(n)$ can be solved from Eqs. (11) and (12). Using
Eqs.(15),(16),(5),and (6), we can obtain the explicit expressions
for $k_{i}(n)$ and $l_{i}(n)$. Eqs.(11) and (12) can be rewritten
as

\begin{equation}\label{17}
  \left(\begin{array}{cc}
     \overline{k^{'}}(n+1) \\
    \overline{l^{'}}(n+1)
\end{array}\right)
=A \left(\begin{array}{cc}
     \overline{k^{'}}(n) \\
    \overline{l^{'}}(n)
\end{array}\right),
\end{equation}
where

\begin{equation}\label{18}
  A=\left(\begin{array}{cc}
   e^{i\gamma}(1-e^{i\beta})W_{k}-e^{i\gamma} & (1-e^{i\beta})W_{l}\\
    e^{i\gamma}(1-e^{i\beta})W_{k} & (1-e^{i\beta})W_{l}-1 \
  \end{array}\right).
\end{equation}
Eq.(17) shows that $\overline{k^{'}}(n)$ and $\overline{l^{'}}(n)$
are dependent on $\gamma$, $\beta$, and $|U_{is}|$. If $U$ is
replaced by a different $U^{\prime}$ where
$|U^{\prime}_{is}|=|U_{is}|$, the analysic forms of
$\overline{k^{'}}(n)$ and $\overline{l^{'}}(n)$ are unaltered. If
$l_{i}(n)$ approaches 0 when $n=n_{0}$, the system lies in

\begin{equation}\label{19}
  |\psi_{t}>=\sum_{i\in M}U_{is}\overline{k^{'}}(n_{0})|i>.
\end{equation}
Generally, the state is not the equally weighted superposition of
marked states. It is related to $U_{is}$. By choosing proper $U$,
some superpositions can be synthesized using generalized Grover's
algorithm. We will solve Eq.(17) in the following section.

\section*{3.Experimental scheme}
   Our experiments use a sample of Carbon-13 labelled chloroform
dissolved in d6-acetone. Data are taken at room temperature with a
Bruker DRX 500 MHz spectrometer. The resonance frequencies
$\nu_{1}=125.76$ MHz for $^{13}C$, and $\nu_{2}=500.13$ MHz for
$^{1}H$. The coupling constant $J$ is measured to be 215 Hz. If
the magnetic field is along $\hat{z}$-axis, and let $\hbar=1$, the
Hamitonian of this system is described by
\begin{equation}\label{20}
  H=-2\pi\nu_{1}I_{z}^{1}-2\pi\nu_{2}I_{z}^{2}+2\pi J I_{z}^{1}
  I_{z}^{2},
\end{equation}
where $I_{z}^{k}(k=1,2)$ are the matrices for $\hat{z}$-component
of the angular momentum of the spins [8]. In the rotating frame of
spin $k$, the evolution caused by a radio-frequency(rf) pulse on
resonance along $\hat{x}$ or $-\hat{y}$-axis is denoted as $
X_{k}(\varphi_{k})= e^{i\varphi_{k}I_{x}^{k}}$ or $
Y_{k}(-\varphi_{k})= e^{-i\varphi_{k}I_{y}^{k}}$, where
$\varphi_{k}=B_{1}\gamma_{k}t_{p}$ with $k$ specifying the
affected spin. $B_{1} $, $\gamma_{k}$ and $t_{p}$ represent the
strength of rf pulse, gyromagnetic ratio and the width of rf
pulse, respectively. The pulse used above is denoted as
$[\varphi]_{x}^{k}$ or $[-\varphi]_{y}^{k}$.
 The coupled-spin evolution is denoted as
\begin{equation}\label{21}
  [t]=e^{-i2\pi J I_{z}^{1} I_{z}^{2}},
\end{equation}
where $t$ is evolution time. The predefined pseudo-pure state
\begin{equation}\label{22}
  |s>=|\uparrow>_{1}|\uparrow>_{2}=\left(\begin{array}{c}
    1 \\
    0 \\
    0 \\
    0 \
  \end{array}\right)
\end{equation}
is prepared by using spatial averaging [9], where $|\uparrow>_{k}$
denotes the state of spin $k$. For convenience, the notation
$|\uparrow>_{1}|\uparrow>_{2}$ is simplified as
$|\uparrow\uparrow>$. The basis states are arrayed as
$|\uparrow\uparrow>,|\uparrow\downarrow>,|\downarrow\uparrow>$, $
|\downarrow\downarrow>$. $U$ is chosen as
$U=Y_{1}(\varphi_{1})Y_{2}(\varphi_{2})$ represented as
\begin{equation}\label{23}
  U=\left(\begin{array}{cccc}
    c_{1}c_{2} & c_{1}s_{2} & s_{1}c_{2} &s_{1}s_{2} \\
    -c_{1}s_{2} &  c_{1}c_{2} & -s_{1}s_{2} & s_{1}c_{2} \\
    -s_{1}c_{2} & -s_{1}s_{2} & c_{1}c_{2} & c_{1}s_{2} \\
    s_{1}s_{2} & -s_{1}c_{2} & -c_{1}s_{2} & c_{1}c_{2} \
  \end{array}\right),
 \end{equation}
where $c_{k}\equiv cos(\varphi_{k}/2)$, $s_{k}\equiv
sin(\varphi_{k}/2)$. When $|\uparrow\uparrow>$ and
$|\downarrow\downarrow>$ are the two marked states, $U$ can be
chosen as $U=Y_{1}(\frac{\pi}{2})Y_{2}(\frac{\pi}{2})$ described
by

\begin{equation}\label{24}
  U=\frac{1}{2}\left(\begin{array}{cccc}
    1 &1 & 1 & 1 \\
    -1& 1 &-1 & 1 \\
   -1 &-1 & 1 & 1 \\
    1 & -1 & -1 & 1 \
  \end{array}\right).
\end{equation}
One can find that $W_{k}=W_{l}=\frac{1}{2}$. When
$\gamma=\beta=-\frac{\pi}{2}$, $I_{t}^{\gamma}$ and
$I_{s}^{\beta}$ can be represented as

\begin{equation}\label{25}
  I_{14}^{-\frac{\pi}{2}}=\left(\begin{array}{cccc}
    -i & 0 & 0 & 0 \\
    0& 1 & 0 & 0 \\
    0 & 0 & 1 & 0 \\
    0 & 0 & 0 & -i \
  \end{array}\right),
 \end{equation}

\begin{equation}\label{26}
  I_{s}^{-\frac{\pi}{2}}=\left(\begin{array}{cccc}
    -i & 0 & 0 & 0 \\
    0& 1 & 0 & 0 \\
    0 & 0 & 1 & 0 \\
    0 & 0 & 0 & 1 \
  \end{array}\right).
 \end{equation}
Using the values of $\beta$, $\gamma$, $W_{k}$, and $W_{l}$,
Eq.(18) is expressed by

\begin{equation}\label{27}
  A=\frac{1}{2}\left(\begin{array}{cc}
    1+i & 1+i\\
    1-i & i-1\
  \end{array}\right)
\end{equation}
In order to obtain explicit expressions for $\overline{k^{'}}(n)$
and $\overline{l^{'}}(n)$, we introduce the diagonal matrix
represented as

\begin{equation}\label{28}
A_{D}=S^{-1}AS\equiv\left(\begin{array}{cc}
    \lambda_{+} & 0\\
    0 & \lambda_{-}\
  \end{array}\right)
\end{equation}
The eigenvalues of matrix $A$ are the solutions of $\det(A-\lambda
I)=0$. They are expressed as $\lambda_{+}=e^{i\frac{\pi}{6}}$ ,
$\lambda_{-}=e^{i\frac{5\pi}{6}}$. $S$ and $S^{-1}$ are expressed
as

\begin{equation}\label{29}
 S=\left(\begin{array}{cc}
    1 & 1\\
    \frac{\sqrt{3}-1}{1+i} & -\frac{\sqrt{3}+1}{1+i}\
  \end{array}\right),
\end{equation}

\begin{equation}\label{30}
S^{-1}=\left(\begin{array}{cc}
    \frac{\sqrt{3}+1}{2\sqrt{3}} & \frac{1+i}{2\sqrt{3}}\\
    \frac{\sqrt{3}-1}{2\sqrt{3}} & -\frac{1+i}{2\sqrt{3}}\
  \end{array}\right).
\end{equation}

The solution of Eq.(17) can be expressed as

\begin{equation}\label{31}
  \left(\begin{array}{cc}
     \overline{k^{'}}(n) \\
    \overline{l^{'}}(n)
\end{array}\right)
=A^{n} \left(\begin{array}{cc}
     \overline{k^{'}}(0) \\
    \overline{l^{'}}(0)
\end{array}\right),
\end{equation}
where $A^{n}=SA_{D}^{n}S^{-1}$, expressed as

\begin{equation}\label{32}
 A^{n}=\frac{1}{2\sqrt{3}}\left(\begin{array}{cc}
    (\sqrt{3}+1)e^{i\frac{n\pi}{6}}+(\sqrt{3}-1)e^{i\frac{5n\pi}{6}}
     & (1+i)(e^{i\frac{n\pi}{6}}-e^{i\frac{5n\pi}{6}})\\
    (1-i)(e^{i\frac{n\pi}{6}}-e^{i\frac{5n\pi}{6}})
     &(\sqrt{3}-1)e^{i\frac{n\pi}{6}}+(\sqrt{3}+1)e^{i\frac{5n\pi}{6}} \
  \end{array}\right).
\end{equation}
When $n=1$, we obtain that
$\overline{k^{'}}(1)=\sqrt{2}e^{i\frac{\pi}{4}}$, and
$\overline{l^{'}}(1)=0$. Using $U_{11}=\frac{1}{2}$,
$U_{41}=\frac{1}{2}$, we obtain that
$k_{1}(1)=e^{i\frac{\pi}{4}}/\sqrt{2}$, and
$k_{4}(1)=e^{i\frac{\pi}{4}}/\sqrt{2}$. The system lies in state

\begin{equation}\label{33}
  |\psi_{1}>=(|\uparrow\uparrow>+|\downarrow\downarrow>)e^{i\frac{\pi}{4}}/\sqrt{2}.
\end{equation}
The overall phase can be ignored.

 If $U$ is chosen as
$U=Y_{1}(-\frac{\pi}{2})Y_{2}(\frac{\pi}{2})$, Eq.(32) is
unaltered. We also obtain
$\overline{k^{'}}(1)=\sqrt{2}e^{i\frac{\pi}{4}}$, and
$\overline{l^{'}}(1)=0$. Noting that $U_{11}=\frac{1}{2}$, and
$U_{41}=-\frac{1}{2}$, we obtain that
$k_{1}(1)=e^{i\frac{\pi}{4}}/\sqrt{2}$, and
$k_{4}(1)=-e^{i\frac{\pi}{4}}/\sqrt{2}$. The system lies state

\begin{equation}\label{34}
  |\psi_{2}>=(|\uparrow\uparrow>-|\downarrow\downarrow>)e^{i\frac{\pi}{4}}/\sqrt{2}.
\end{equation}

Considering the experimental convenience, if the marked states are
$|\uparrow\downarrow>$ and $|\downarrow\uparrow>$, we choose
$I_{t}^{\gamma}$ as $I_{23}^{\gamma}=iI_{14}^{-\frac{\pi}{2}}$,
where $\gamma=\frac{\pi}{2}$. In matrix notation,
$I_{23}^{\gamma}$ is represented as

\begin{equation}\label{35}
  I_{23}^{\frac{\pi}{2}}=\left(\begin{array}{cccc}
    1 & 0 & 0 & 0 \\
    0& i & 0 & 0 \\
    0 & 0 & i & 0 \\
    0 & 0 & 0 & 1 \
  \end{array}\right).
 \end{equation}
$\beta$ is changed to $\frac{\pi}{2}$ to satisfy phase matching
[10]. $I_{s}^{\frac{\pi}{2}}$ is described by

\begin{equation}\label{36}
  I_{s}^{\frac{\pi}{2}}=\left(\begin{array}{cccc}
    i & 0 & 0 & 0 \\
    0& 1 & 0 & 0 \\
    0 & 0 & 1 & 0 \\
    0 & 0 & 0 & 1 \
  \end{array}\right).
 \end{equation}
 When $\gamma=\beta=\frac{\pi}{2}$,
and $U=Y_{1}(\frac{\pi}{2})Y_{2}(\frac{\pi}{2})$, or
$U=Y_{1}(-\frac{\pi}{2})Y_{2}(\frac{\pi}{2})$, the solution of
Eq.(17) can be obtained by replacing $i$ in Eqs.(27)-(32) by $-i$.
We obtain $\overline{k^{'}}(1)=\sqrt{2}e^{-i\frac{\pi}{4}}$, and
$\overline{l^{'}}(1)=0$. When
$U=Y_{1}(\frac{\pi}{2})Y_{2}(\frac{\pi}{2})$, we obtain
$k_{2}(1)=-e^{-i\frac{\pi}{4}}/\sqrt{2}$, and
$k_{3}(1)=-e^{-i\frac{\pi}{4}}/\sqrt{2}$, using
$U_{21}=-\frac{1}{2}$, and $U_{31}=-\frac{1}{2}$. The system lies
in state

\begin{equation}\label{37}
  |\psi_{3}>=-(|\uparrow\downarrow>+|\downarrow\uparrow>)e^{-i\frac{\pi}{4}}/\sqrt{2}.
\end{equation}
Similarly, when $U=Y_{1}(-\frac{\pi}{2})Y_{2}(\frac{\pi}{2})$, we
obtain $k_{2}(1)=-e^{-i\frac{\pi}{4}}/\sqrt{2}$, and
$k_{3}(1)=e^{-i\frac{\pi}{4}}/\sqrt{2}$, using
$U_{21}=-\frac{1}{2}$, and $U_{31}=\frac{1}{2}$. The system lies
in state

\begin{equation}\label{38}
  |\psi_{4}>=-(|\uparrow\downarrow>-|\downarrow\uparrow>)e^{-i\frac{\pi}{4}}/\sqrt{2}.
\end{equation}
$|\psi_{1}>$, $|\psi_{2}>$, $|\psi_{3}>$ and $|\psi_{4}>$ are the
four EPR states. They are very useful in quantum information and
have been implemented in experiments [11][12]. Based on the
discussion above, they can be synthesized by generalized Grover's
algorithm. Other entangled states can be obtained by choosing
other $U$. For example, if $U$ is chosen as
\begin{equation}\label{39}
 U=X_{1}(\frac{\pi}{2})Y_{2}(\frac{\pi}{2})=
 \frac{1}{2}\left(\begin{array}{cccc}
    1 &1 & i & i \\
    -1& 1 &-i & i \\
   i &i & 1 & 1 \\
    -i & i & -1 & 1 \
  \end{array}\right),
\end{equation}
and $\gamma=\beta=-\frac{\pi}{2}$, entangled state
$(|\uparrow\uparrow>-i|\downarrow\downarrow>)e^{i\frac{\pi}{4}}/\sqrt{2}$
is obtained after one iteration. The target states, such as
$|\psi_{1}>$ and $|\psi_{2}>$, can also be obtained by matrix
multiplication. If replacing $n$ in Eq.(32) by $n+3$, one finds
$A^{n+3}=iA^{n}$. This fact shows that $\overline{k^{'}}(n)$ and
 $\overline{l^{'}}(n)$ both have a period of 3.

\section*{4.Experimental procedure}
   The equilibrium density matrix can be represented as

\begin{equation}\label{40}
  \rho_{eq}=\gamma_{1}I_{z}^{1}+\gamma_{2}I_{z}^{2}.
\end{equation}
The rf and gradient pulse sequence
$[\alpha]_{x}^{2}-[grad]_{z}-[\pi/4]_{x}^{1}-1/4J-[\pi]_{x}^{1,2}-1/4J-[-\pi]_{x}^{1,2}
-[-\pi/4]_{y}^{1}-[grad]_{z}$ transforms the system from the
equilibrium state into the state represented as

\begin{equation}\label{41}
  \rho_{s}=I_{z}^{1}/2+I_{z}^{2}/2+I_{z}^{1}I_{z}^{2}=\frac{1}{4}\left(\begin{array}{cccc}
    3 & 0 & 0 & 0 \\
    0 & -1 & 0 & 0 \\
    0 & 0 & -1 & 0 \\
    0 & 0 & 0 & -1 \
  \end{array}\right),
\end{equation}
which can be used as the pseudo-pure state $|\uparrow\uparrow>$
[13]. $\alpha=\arccos(\gamma_{1}/2\gamma_{2})$,
 $[grad]_{z}$ denotes gradient pulse along $\hat{z}$-axis, and the
symbol 1/4J means that the system evolutes under $H$ described as
Eq.(20) for 1/4J time when pulses are closed. The pulses are
applied from left to right. $[\pi]_{x}^{1,2}$ denotes a
nonselective pulse (hard pulse). The evolution caused by the pulse
sequence $1/4J-[\pi]_{x}^{1,2}-1/4J-[-\pi]_{x}^{1,2}$ is
equivalent to the coupled-spin evolution $[1/2J]$ described in
Eq.(21) [14]. $\pi$ pulses are applied in pairs each of which take
opposite phases in order to reduce the error accumulation caused
by imperfect calibration of $[\pi]$ pulses [15]. $U$ is realized
by $[\pm\pi/2]_{y}^{1}-[\pi/2]_{y}^{2}$, corresponding to
$Y_{1}(\pm\frac{\pi}{2})Y_{2}(\frac{\pi}{2})$, respectively.
Because $I_{23}^{\frac{\pi}{2}}=iI_{14}^{-\frac{\pi}{2}}$,
$I_{23}^{\frac{\pi}{2}}$ and $I_{14}^{-\frac{\pi}{2}}$ can be
realized by the same sequence
$1/4J-[\pi]_{x}^{1,2}-1/4J-[-\pi]_{x}^{1,2}$. By modifying the
pulses used in Refs.[6][16], we realize $I_{s}^{-\frac{\pi}{2}}$
by $1/8J-[\pi]_{x}^{1,2}-1/8J-[-\pi]_{x}^{1,2}-[-\pi/2]_{y}^{1,2}
-[-\pi/4]_{x}^{1,2}-[\pi/2]_{y}^{1,2}$, and
$I_{s}^{\frac{\pi}{2}}$ by
$15/8J-[\pi]_{x}^{1,2}-15/8J-[-\pi]_{x}^{1,2}-[-\pi/2]_{y}^{1,2}-[\pi/4]_{x}^{1,2}-
[\pi/2]_{y}^{1,2}$. When
$U=Y_{1}(\frac{\pi}{2})Y_{2}(\frac{\pi}{2})$,
$(-UI_{s}^{-\frac{\pi}{2}}U^{\dag}I_{14}^{-\frac{\pi}{2}})U$
transforms the pseudo-pure state $|\uparrow\uparrow>$ into state
$|\psi_{1}>$, and
$(-UI_{s}^{\frac{\pi}{2}}U^{\dag}I_{23}^{\frac{\pi}{2}})U$
transforms $|\uparrow\uparrow>$ into $|\psi_{3}>$, where $U$
transforms $|\uparrow\uparrow>$ into the initial state $|g(0)>$,
and $( )$ indicates Grover iteration. When
$U=Y_{1}(-\frac{\pi}{2})Y_{2}(\frac{\pi}{2})$,
$(-UI_{s}^{-\frac{\pi}{2}}U^{\dag}I_{14}^{-\frac{\pi}{2}})U$
transforms $|\uparrow\uparrow>$ into $|\psi_{2}>$, and
$(-UI_{s}^{\frac{\pi}{2}}U^{\dag}I_{23}^{\frac{\pi}{2}})U$
transforms $|\uparrow\uparrow>$ into $|\psi_{4}>$. The results are
expressed by density matrixes. For example, the density matrix
corresponding to $|\psi_{1}>$ is represented as

\begin{equation}\label{42}
  \rho_{1}=(I_{x}^{1}I_{x}^{2}-I_{y}^{1}I_{y}^{2}+I_{z}^{1}I_{z}^{2})
  =\left(\begin{array}{cccc}
    0.25 & 0 & 0 & 0.5 \\
    0 & -0.25 & 0& 0 \\
    0 & 0 & -0.25 & 0 \\
    0.5 & 0 & 0 & 0.25 \
  \end{array}\right),
\end{equation}
which is equivalent to $|\psi_{1}><\psi_{1}|$. A readout pulse
$[\pi/2]_{y}^{2}$ transforms $\rho_{1}$ into $\rho_{1r}$
represented as

\begin{equation}\label{43}
  \rho_{1r}=\frac{1}{4}\left(\begin{array}{cccc}
    0 & -1 & 1 & 1 \\
    -1 & 0 & -1 & -1 \\
    1 & -1 & 0 & 1 \\
    1 & -1 & 1 & 0 \
  \end{array}\right).
\end{equation}
The information on matrix elements (1,3) and (2,4) in Eq.(43) can
be directly obtained in the carbon spectrum, and the information
on elements (1,2) and (3,4) can be directly obtained in the proton
spectrum. Similarly, when the system lies in $|\psi_{2}>$,
$|\psi_{3}>$, or $|\psi_{4}>$, the readout pulse $[\pi/2]_{y}^{2}$
transforms the system into the state represented as
\begin{equation}\label{44}
  \rho_{2r}=\frac{1}{4}\left(\begin{array}{cccc}
    0 & -1 & -1 & -1 \\
    -1 & 0 & 1 & 1 \\
    -1 & 1 & 0 & 1 \\
    -1 & 1 & 1 & 0 \
  \end{array}\right),
\end{equation}

\begin{equation}\label{45}
  \rho_{3r}=\frac{1}{4}\left(\begin{array}{cccc}
    0 & 1 & 1 & -1 \\
    1 & 0 & 1 & -1 \\
    1 & 1 & 0 & -1 \\
    -1 & -1 & -1 & 0 \
  \end{array}\right),
\end{equation}
or
\begin{equation}\label{46}
  \rho_{4r}=\frac{1}{4}\left(\begin{array}{cccc}
    0 & 1 & -1 & 1\\
    1 & 0 & -1 & 1 \\
    -1 & -1 & 0 &-1 \\
    1 & 1 & -1 & 0 \
  \end{array}\right).
\end{equation}
Through observing the matrix elements (1,3), (2,4), (1,2) and
(3,4) in Eqs.(43)-(46), one can distinguish the four EPR states.

\section*{5.Results}
  In experiments, for each target state, the carbon spectrum and
proton spectrum are recorded in two experiments. For different
target states, carbon spectra or proton spectra are recorded in an
identical fashion. Because the absolute phase of an NMR signal is
not meaningful, we must use reference signals to adjust carbon
spectra and proton spectra so that the relative phases of the
signals are meaningful [17]. When the system lies in the
pseudo-pure state described as Eq.(41), the readout pulses
$[\pi/2]_{y}^{1}$ and $[\pi/2]_{y}^{2}$ transform it into states
represented as
\begin{equation}\label{47}
  \rho_{sr1}=\frac{1}{4}\left(\begin{array}{cccc}
    1 & 0 & -2 & 0\\
    0 & -1 & 0 & 0 \\
    -2 & 0 & 1 & 0 \\
    0 & 0 & 0 & -1 \
  \end{array}\right),
\end{equation}
and
\begin{equation}\label{48}
  \rho_{sr2}=\frac{1}{4}\left(\begin{array}{cccc}
    1 & -2 & 0 & 0\\
    -2 & 1 & 0 & 0 \\
    0 & 0 & -1 & 0 \\
    0 & 0 & 0 & -1 \
  \end{array}\right),
\end{equation}
respectively. In the carbon spectrum or proton spectrum, there is
only one MNR peak corresponding to element (1,3) in $\rho_{sr1}$
or to element (1,2) in $\rho_{sr2}$. Through calibrating the
phases of the two signals, the two peaks are adjusted into
absorbtion shapes which are shown as Fig.1(a)a for carbon spectrum
and Fig.1(b) for proton spectrum. The two signals are used as
reference signals of which phases are recorded to calibrate the
phases of signals in other carbon spectra and proton spectra,
respectively. One should note that the minus elements in Eq.(47)
and Eq.(48) are corresponding to the positive peaks in Fig.1(a)
and Fig.1(b).

  We implement generalized Grover's algorithm starting with the
initial state $|g(0)>$. $GI_{t}^{\gamma}$ transforms $|g(0)>$ into
one of EPR states. If no readout pulse is applied, the amplitudes
of peaks is so small that they can be ignored. By applying the
spin-selective readout pulse $[\pi/2]_{y}^{2}$, we obtain the
carbon spectra as shown in Figs.2(a), (b), (c), and (d), and the
proton spectra as shown in Figs.3(a), (b), (c), and (d). Fig.2(a)
and Fig.3(a) are corresponding to $|\psi_{1}>$, Fig.2(b) and
Fig.3(b) to $|\psi_{2}>$, Fig.2(c) and Fig.3(c) to $|\psi_{3}>$,
and Fig.2(d) and  Fig.3(d) to $|\psi_{4}>$. In Fig.2(a), for
example, the right and left peaks are corresponding to the matrix
elements (1,3) and (2,4) in Eq.(43), respectively. Similarly, in
Fig.3(a), the two peaks are corresponding to the matrix elements
(1,2) and (3,4) in Eq.(43). The phases of the signals corroborate
the synthesis of EPR states.

\section*{6.Conclusion}
  In experiments, we demonstrate generalized Grover's algorithm
of 2 marked states. The results show that generalized Grover's
algorithm is efficient for the case of $N/2$ marked states. The
original Grover's algorithm, however, does not work for this case.
EPR states are synthesized using the algorithm. For the case of
multiple marked states, the final signal is an average over all
the marked states. It is difficult or impossible to deduce
anything about individual marked states from the ensemble average
[18]. In our work, however, the generalized Grover's algorithm is
viewed as a technique for synthesizing a particular kind of
superposition of marked states [5]. The superposition is described
as a density matrix of which elements can be obtained in the NMR
spectrum by readout pulses [19].
\section*{Acknowledgment}
  This work was partly supported by the National Nature Science
Foundation of China. We are also grateful to Professor Shouyong
Pei of Beijing Normal University for his helpful discussions on
the principle of quantum algorithm.
\newpage
\bibliographystyle{article}

\newpage
{\begin{center}\large{Figure Captions}\end{center}
\begin{enumerate}
\item The carbon spectrum (Fig.1(a)) obtained through selective readout
pulse for $^{13}C$ $[\pi/2]_{y}^{1}$  and the proton spectrum
(Fig.2(b)) obtained through selective readout pulse for $^{1}H$
$[\pi/2]_{y}^{2}$ when the two-spin system lies in pseudo-pure
state $|\uparrow\uparrow>$. The two peaks are adjusted into
absorbtion shapes. The two signals are used as reference signals
to adjust other spectra.
\item Carbon spectra obtained through $[\pi/2]_{y}^{2}$ after EPR
 states are synthesized. Figs.2(a), (b), (c) and (d) are
corresponding to states
$(|\uparrow\uparrow>+|\downarrow\downarrow>)/\sqrt{2}$,
$(|\uparrow\uparrow>-|\downarrow\downarrow>)/\sqrt{2}$,
$(|\uparrow\downarrow>+|\downarrow\uparrow>)/\sqrt{2}$, and
$(|\uparrow\downarrow>-|\downarrow\uparrow>)/\sqrt{2}$,
respectively.
\item Proton spectra obtained through $[\pi/2]_{y}^{2}$ after EPR
 states are synthesized. Figs.3(a), (b), (c), and (d) are
 corresponding to states
$(|\uparrow\uparrow>+|\downarrow\downarrow>)/\sqrt{2}$,
$(|\uparrow\uparrow>-|\downarrow\downarrow>)/\sqrt{2}$,
$(|\uparrow\downarrow>+|\downarrow\uparrow>)/\sqrt{2}$, and
$(|\uparrow\downarrow>-|\downarrow\uparrow>)/\sqrt{2}$,
respectively.
\end{enumerate}
\begin{figure}{1}
\includegraphics[]{figg.1.eps}
\caption{}
\end{figure}
\begin{figure}{2}
\includegraphics[]{figg.2.eps}
\caption{}
\end{figure}
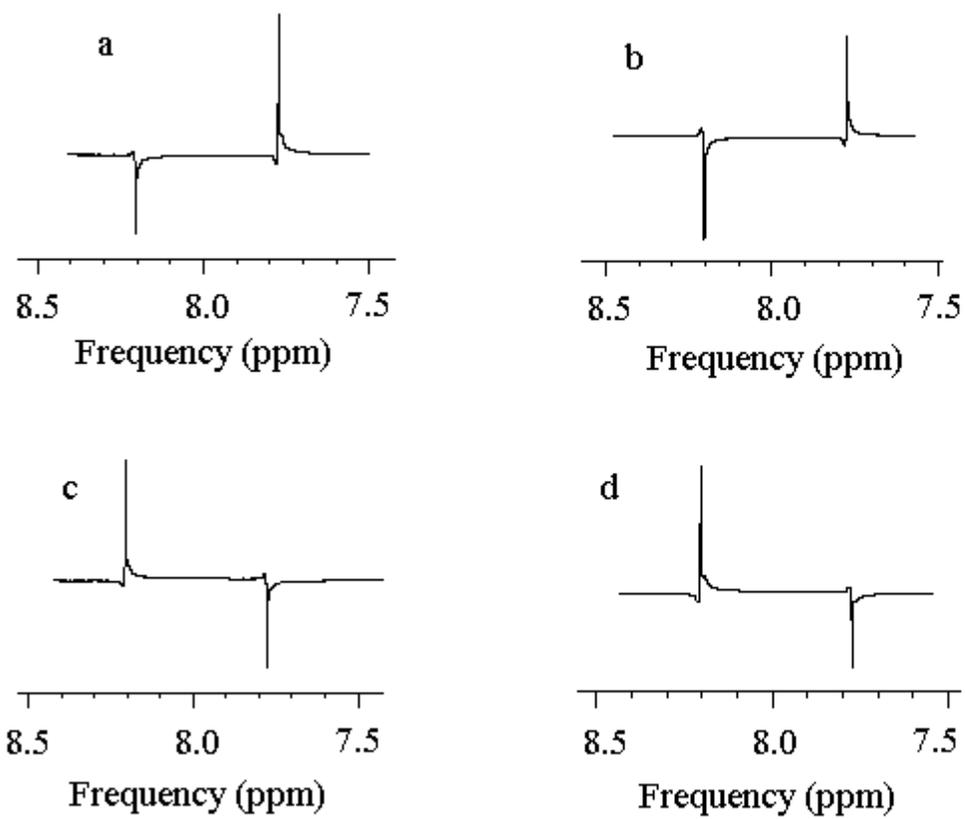
\begin{figure}{3}
\includegraphics[]{figg.3.eps}
\caption{}
\end{figure}

\end{document}